\documentstyle[11pt,newpasp,twoside]{article}
\input{epsf}
\begin{document}
\title{Rapidly rotating strange stars} \author{Gondek-Rosi\'nska D., Haensel P.,
  Zdunik J. L.}  \affil{Copernicus Astronomical Center, Bartycka 18, 
Warszawa, Poland}
\author{Gourgoulhon E.}  \affil{DARC, CNRS, Observatoire
  de Paris, F-92195 Meudon Cedex, France}

\begin{abstract}
  We study effects of the strange quark mass, $m_s$, and of the QCD
  interactions, calculated to lowest order in $\alpha_{\rm c}$, on the
  rapid rotation of strange stars (SS). The influence of rotation on global
  parameters of SS is greater than in the case of the
  neutron stars (NS). We show that independently of $m_s$ and $\alpha_{\rm c}$
  the ratio of the rotational kinetic energy to the absolute value of
  the gravitational potential energy $T/W$ for a rotating SS
  is significantly higher than for an ordinary NS. This
  might indicate that rapidly rotating SS could be important sources of
  gravitational waves.
\end{abstract}
%\section{Results}
Recently, it has been shown that SS are not a subject to the r-mode
instability (Madsen 1998, Stergioulas 1998) in a contrast to newly born
NS.  Therefore SS formed in a collapse of rotating stellar cores can
rotate very fast. We have calculated exact models of the uniformly
rotating SS using multi-domain spectral methods, used previously for
calculating rapidly rotating SS models with $m_s=\alpha_{\rm c}=0$
(Gourgoulhon et al.,1999, A\&A 349, 851). We construct $M-R$ curves for 
SSs rotating with a broad range of angular velocity and calculate the ratio $T/W$,
relevant for stars stability with respect to triaxial deformations. We
present results for two EOS of strange quark matter, both with Bag
Constant $B=60~{\rm MeV/fm^3}$: SQ1 with $m_s=200~{\rm MeV/c^2}$~,
$\alpha_{\rm c}=0$, and SQ2 with $m_s=200~{\rm MeV/c^2}$, $\alpha_{\rm
  c}=0.6$ and compared them with SQ0 EOS with $m_s=\alpha_{\rm c}=0$.
Results for different values of $B$ can be obtained from those for
$B=60~{\rm MeV/fm^3}$ using scaling properties of corresponding
quantities with respect to the change of $B$.  The main conclusions can be
summarised as follows:

 $\bullet$ There is a large increase of the maximum mass and corresponding
  radius of SS as the rotation rate increases from zero to the
  equatorian mass-shedding limit (Fig. 1a). The rotation increases
  the maximum allowable mass of SS by $\sim 40\%$ compared to $\sim 20\%$
  for NS and the equatorial radius by $\sim 50\%$ compared to $\sim
  30\%$ for NS. At fixed $B$ the maximum mass is for SQ0 model
  rotating with Keplerian velocity (the maximum mass $M=2.861
  ~M_\odot$ is for the lowest allowable $B=58.9~{\rm MeV/fm^3}$).
  
  $\bullet$ At fixed $B$, the effect of $m_s$ on rotating SS models is
  much larger than that of $\alpha_{\rm c}$. Both decreases stellar
  mass and radius (Fig. 1a, 1b). The M-R curves for sequences of SS
  rotating with high fixed velocity $f=\Omega/2\pi$ reminds M(R)
  relations for ordinary NS. When $f$ increases the $M(R)$ curves for
  all models are more and more flat. Fast rotation causes a large
  increase of a radius of SS especially for small masses.
  
 $\bullet$ The values of $T/W$ for SS rotating at the Keplerian velocity
  can significantly exceed 0.2 (see Fig. 1c for supramassive stars).
  There are much greater than for NS (more that twice). All this seems
  to indicate that triaxial instabilities could develop more easily in
  rotating SS than in NS. The value of $T/W$ increases with decreasing
  stellar mass for all considered models of SS. We have $M=2.82,\ 
  2.53,\ 2.52,\ M_\odot$ for $T/W=0.2$ and $M=2.14,\ 2.01,\ 1.99\ 
  M_\odot$ for $T/W=0.25$ for SQ0, SQ1, SQ2 respectively.
  
 $\bullet$ Unusually high ratio $T/W$ is characteristic not exclusively
  for supramassive stars but can be reached also for
  rapidly rotating SS with small and moderate masses (Fig. 1d). For
  stars rotating with fixed frequency, $T/W$ increases with decreasing
  central density. The large value of $T/W$ results from a flat density
  profile combined with strong equatorial flattening of rapidly
  rotating SS.

%##############################
\begin{figure}
\begin{center}
\leavevmode
\epsfxsize=10cm
\epsfysize=9cm
%\plotone{win.eps}
\epsfbox{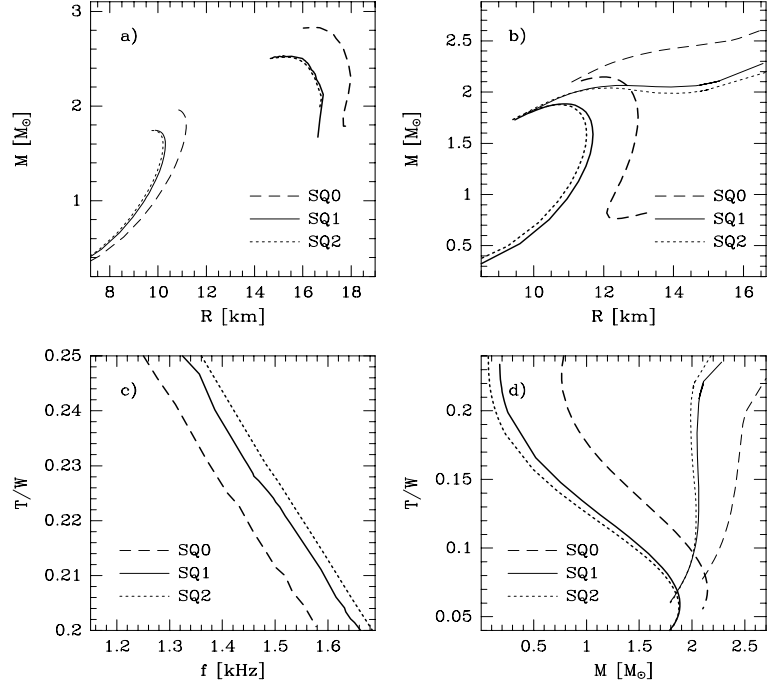}
\end{center}
\caption{a) Mass vs. radius for maximally rotating 
  (thick lines) and static (thin lines) SS. b) Mass vs. radius for
  sequences of SS rotating with frequency $f=1.1\ {\rm
    kHz}$ (thick lines) and $f = 1.4\ {\rm kHz}$ (thin lines). c)
  Kinetic to gravitational energy ratio vs. the rotation
  frequency for supramassive SS rotating with Keplerian
  velocity. d) $T/W$ vs. mass for SS rotating with fixed $f = 1.1\ 
  {\rm and}\ 1.4\ {\rm kHz}$ (thick and thin lines, respectively). The ratio 
$T/W$ increases along each sequence as central density decreases.}
\end{figure}

\acknowledgments 
This work was supported in part by the following
grants KBN-2P03D01413 and KBN-2P03D02117.

%\begin{references}
%\reference Gourgoulhon, E, Haensel P, Livine, R. et al., A\&A, 1999
%\reference Madsen J., 1998, Phys. Rev. Lett. 81, 3311
%\reference Stergioulas N., 1998, Living Reviews in Relativity 1
%\end{reference}

\end{document}